\pdfoutput=1 

\documentclass[11pt,letterpaper,twocolumn]{scrartcl}

\makeatletter
\DeclareOldFontCommand{\rm}{\normalfont\rmfamily}{\mathrm}
\DeclareOldFontCommand{\sf}{\normalfont\sffamily}{\mathsf}
\DeclareOldFontCommand{\tt}{\normalfont\ttfamily}{\mathtt}
\DeclareOldFontCommand{\bf}{\normalfont\bfseries}{\mathbf}
\DeclareOldFontCommand{\it}{\normalfont\itshape}{\mathit}
\DeclareOldFontCommand{\sl}{\normalfont\slshape}{\@nomath\sl}
\DeclareOldFontCommand{\sc}{\normalfont\scshape}{\@nomath\sc}
\makeatother

\usepackage[top=2.4cm, bottom=2.4cm, left=1.5cm, right=1.5cm,footskip=0.8cm]{geometry}

\usepackage[T1]{fontenc}
\usepackage[utf8]{inputenc}
\usepackage{abstract}

\usepackage[hidelinks]{hyperref}
\usepackage[parfill]{parskip}
\usepackage{slashed}
\usepackage{microtype}
\usepackage{csquotes}
\usepackage{graphicx}
\usepackage{amsmath}
\usepackage{amssymb}
\usepackage{braket}
\usepackage{booktabs}
\usepackage{caption}
\usepackage{subcaption}
\usepackage{todonotes}
\usepackage{grffile}

\usepackage[numbers,sort&compress]{natbib}

\usepackage{multirow}

\usepackage{fancyhdr}
\usepackage[yyyymmdd,hhmmss]{datetime}
\fancypagestyle{plain}
{
}
\usepackage{cleveref}

\usepackage[auth-sc,affil-it]{authblk}

\usepackage{scalefnt}
\newcommand{\abbrev}{\scalefont{.9}}

\newcommand{\NLL}{\text{\abbrev NLL}}
\newcommand{\NNLL}{\text{\abbrev N$^2$LL}}
\newcommand{\NNNLL}{\text{\abbrev N$^3$LL}}
\newcommand{\NNLO}{\text{\abbrev NNLO}}
\newcommand{\NNNLO}{\text{\abbrev N$^3$LO}}
\newcommand{\NLO}{\text{\abbrev NLO}}

\newcommand{\LO}{\text{\abbrev LO}}

\newcommand{\QCD}{\text{\abbrev QCD}}

\newcommand{\PDF}{\text{\abbrev PDF}}

\newcommand{\LHC}{\text{\abbrev LHC}}

\newcommand{\CMS}{\text{\abbrev CMS}}
\newcommand{\ATLAS}{\text{\abbrev ATLAS}}

\newcommand{\SCET}{\text{\abbrev SCET}}
\newcommand{\RGE}{\text{\abbrev RGE}}

\newcommand{\MCFM}{\text{\abbrev MCFM}}

\newcommand{\CuTeMCFM}{\texttt{CuTe-MCFM}}

\newcommand{\TeV}{\text{TeV}}
\newcommand{\GeV}{\text{GeV}}
\newcommand{\SI}[2]{ \text{$#1$}\,#2 }

\newcommand{\abs}[1]{\lvert#1\rvert}

\newcounter{notecount}

\makeatletter
\renewcommand\maketitle{
	\begin{center}
		{\huge\bfseries\@title\par\vspace{0.3em}}
		{\scshape\@author, \@date}
	\end{center}
}
\makeatother

\fancypagestyle{firstpage}{%
	\lhead{}
	\rhead{}
}

\begin{document}
	
	\thispagestyle{firstpage}
	\title{\LARGE { The Diphoton $q_T$ spectrum at N$^3$LL$^\prime$+NNLO }}
	
	\author[1]{Tobias Neumann}
	
	\affil[1]{Department of Physics, Brookhaven National Laboratory, Upton, New York 11973, USA}
	
	\date{}
	\twocolumn[
	\maketitle
	
	\vspace{0.5cm}
	
	\begin{onecolabstract}
		\vspace{0.5cm} 
		We present a $q_T$-resummed calculation of diphoton production at order {\abbrev 
		N$^3$LL$^\prime$+NNLO}. To 
		reach the primed level of accuracy we have implemented the recently published three-loop 
		$\mathcal{O}(\alpha_s^3)$ 
		virtual corrections in the $q\bar{q}$ channel and the three-loop transverse momentum 
		dependent beam functions and 
		combined them with the existing infrastructure of \CuTeMCFM{}, a code performing 		
		resummation at order \NNNLL{}. While the primed
		predictions are parametrically not more accurate, one typically observes from lower orders 
		and other processes that they are the 
		dominant effect of the next order. We include in both the $q\bar{q}$ and 
		loop-induced $gg$ channel the
		hard contributions consistently together at order  $\alpha_s^3$ and find that the 
		resummed $q\bar{q}$ channel without matching stabilizes indeed. Due to large matching 
		corrections and large contributions and uncertainties from the $gg$ channel, the 
		overall improvements are small though. We furthermore study the effect of hybrid-cone photon
		isolation and hard-scale choice on our fully matched results to describe the \ATLAS{} 
		\SI{8}{\TeV} data and find that the hybrid-cone isolation worsens agreement at 
		small $q_T$ compared to smooth-cone isolation.
			\vspace{0.5cm}
	\end{onecolabstract}
	]
	
	\vspace{-5mm}
	
	
	
	The production of prompt isolated photon pairs at hadron colliders is a test of \QCD{} with a 
	clean experimental signature and constitutes as a background to the $h\to\gamma\gamma$ 
	decay. Differential measurements at the \LHC{} are available at \SI{7}{\TeV}, 
	both by \ATLAS{} and \CMS{} \cite{Chatrchyan:2014fsa,Aad:2011mh,ATLAS:2012fgo}, at 
	\SI{8}{\TeV} by ATLAS \cite{ATLAS:2017cvh}, and very recently at \SI{13}{\TeV} also by \ATLAS{} 
	\cite{ATLAS:2021mbt}. Further, many models of new physics predict 
	resonant large mass diphoton decays for which events with diphoton invariant masses of up to 
	$\sim\SI{2}{\TeV}$ \cite{ATLAS:2017ayi,CMS:2018dqv} are used to constrain them.
	
	Photons can either be produced directly or through the fragmentation of \QCD{} 
	partons. Photons produced through fragmentation require treatment of their singular collinear 
	splitting. These singularities can be renormalized into non-perturbative fragmentation 
	functions, but which are available only at \NLO{} so far with large uncertainties 
	\cite{Kaufmann:2017lsd,Gehrmann-DeRidder:1997fom}. In 
	higher-order calculations the collinear singularity is typically removed with a smooth-cone 
	isolation 
	prescription \cite{Frixione:1998jh}, which eliminates the collinear singularity, but does not 
	prevent the usual cancellation of soft singularities. Quite some attention has been paid to the 
	smooth-cone photon isolation procedure in 
	recent years due to the uncertainty associated with the 
	difference to experimental photon isolation \cite{Gehrmann:2020oec,Chen:2019zmr,Catani:2018krb}.
	
	\NNLO{} calculations \cite{Catani:2011qz,Campbell:2016yrh} for diphoton production have shown 
	large perturbative corrections with uncertainties that significantly underestimate the 
	difference going from \NLO{} to \NNLO{}, estimated by using typical scale-variation procedures. 
	This is in 
	part due to a new loop-induced $gg$ 
	channel entering at \NNLO{} \cite{Campbell:2016yrh,Campbell:2011bn,Bern:2001df,Glover:2003cm}, 
	but even true for just the Born-level induced $q\bar{q}$ channel. This 
	has theorists led to deviate from the usual scale-variation prescription and suggest taking 
	(half) the difference between \NLO{} and \NNLO{} as an estimate of the \NNLO{} uncertainty 
	\cite{Catani:2018krb}. This in turn mandates the calculation of higher-order predictions until
	uncertainties obtained from scale-variation stabilize. The inclusion of the
	recently published 	three-loop $q\bar{q}\to\gamma\gamma$ hard function \cite{Caola:2020dfu} in 
	the calculation presented in this paper contributes to these improvements.

Resummed predictions at small $q_T$ are 
available at \NNNLL{} \cite{Becher:2020ugp} and at \NNLL{} \cite{Cieri:2015rqa}, both matched to 
$\NNLO{}_0$ fixed-order and have been computed previously at lower order matched to $\NLO{}_0$ 
\cite{Balazs:2006cc,Balazs:2007hr,Nadolsky:2007ba,Coradeschi:2017zzw}.\footnote{With the 
subscript $0$ we denote that the order is with respect to the Born-level topology $\gamma\gamma$ 
and not with respect to $\gamma\gamma$+jet, i.e. at large $q_T$. Latter would be denoted with a $1$ 
subscript. Therefore \NNNLL{}+\NNLO$_0$ is equivalent to \NNNLL{}+\NLO{}$_1$. This is often 
ambiguous in the literature. In the following we will omit the $0$ subscript.} Recently, a matching 
to 
parton shower has been considered \cite{Alioli:2020qrd}. \NLO{} electroweak effects 
have been studied in ref.~\cite{Chiesa:2017gqx}, which are found to be less than one percent for 
the $q_T$ distribution below \SI{200}{\GeV}, so 
they are not relevant for the results presented in this paper.

Uncertainties in the measured diphoton transverse momentum ($q_T$)
and $\phi^*$ \cite{Banfi:2010cf} distributions at \SI{7}{\TeV} and \SI{8}{\TeV} are about 
10\%. This is about the uncertainty estimated by scale variation from fixed-order \NNLO{} 
($\mathcal{O}(\alpha_s^2)$) and N$^3$LL $q_T$-resummed 
calculations. While the uncertainties seem large, the differences between theory and measurement 
are found to be larger. Tensions start at $q_T\sim\SI{15}{\GeV}$ and rise to differences of 30\% 
between central values for $q_T$ larger than \SI{50}{\GeV} \cite{Becher:2020ugp}.

Before moving on to a discussion about the photon isolation, we need to set up the relevant 
notation.
The smooth-cone photon isolation prescription in this paper \cite{Frixione:1998jh} restricts the 
transverse hadronic 
energy $E_T^\text{had}$ around photons to be 
\begin{eqnarray*}
	E_T^\text{had} \le E_T^\text{iso} \chi^\text{smooth}(r,R_s)\,, \quad \forall r\le R_s, \\
	\chi^\text{smooth}(r,R_s) = \left( \frac{1-\cos(r)}{1-\cos(R_s)} \right)^n\,,
\end{eqnarray*}
where $E_T^\text{iso}$ is an isolation cone energy that can either be fixed or dependent on the 
photon transverse momentum, $R_s$ is the isolation cone radius and $n$ is a parameter.
In this paper we also consider a simple hybrid-cone isolation that takes the smooth-cone isolation 
within an inner radius $R_s$ and a fixed-cone prescription in the outer cone with radius $R_o$:
\begin{eqnarray*}
	\chi^\text{hybrid}(r,R_s,R_o) = \begin{cases}
		\chi^\text{smooth}(r,R_s) & r \le R_s \\
		1 & R_s < r < R_o
	\end{cases}\,.
\end{eqnarray*}

Recently it has been argued that a bulk of the data-theory tensions are an artifact of two effects 
\cite{Gehrmann:2020oec}. The first one is regarding the hard renormalization scale choice of 
$m_{\gamma\gamma}$ that has 
often been used for predictions. Since the diphoton pair is not produced resonantly, 
$m_{\gamma\gamma}$ is not the definite obvious choice for capturing the hard process kinematic 
scale. An 
alternative studied is to take the 
arithmetic 
mean 
of 
the photon transverse momenta $\langle q_T^\gamma\rangle$. The second effect is due to the photon 
isolation, which 
likely needs to be assigned larger uncertainties than previously thought.
The suggestion is to use the hybrid-cone isolation, which allows for a better matching to the 
experimentally used fixed-cone prescription by adjusting the inner cone radius. An isolation 
uncertainty can then be obtained by varying the inner cone radius by some amount.

In ref.~\cite{Gehrmann:2020oec} it is further argued that isolation scheme and hard scale have 
compensating effects, at least for distributions sensitive to the photon separation $\Delta 
R_{\gamma\gamma}$ like the 
$m_{\gamma\gamma}$ distribution.
The authors show that $\mu=m_{\gamma\gamma}$ plus smooth-cone isolation and $\mu=\langle 
q_T^\gamma\rangle$ plus hybrid-cone
isolation should go together, respectively, as there are compensating effects in the low and large 
mass region.
Indeed they show that the agreement between data and theory is best for the combination of $\langle 
q_T^\gamma\rangle$ plus hybrid-cone isolation, but the experimental uncertainties in the large and 
small 
mass regions are also largest, leaving an unclear picture.

Every isolation prescription introduces an unphysical discontinuity due to the presence of step 
functions in the measurement function \cite{Catani:1997xc,Binoth:1999qq}. At \NNLO{} the 
discontinuity translates into a Sudakov singularity that has been studied in more detail in 
ref.~\cite{Gehrmann:2020oec}:
 In the case of the smooth-cone isolation the affected observable is not of 
experimental relevance. But the hybrid-cone isolation places this singularity directly into the 
$q_T^{\gamma\gamma}$ distribution around $E_T^\text{iso}$ and 
its only cure is a sufficiently large experimental 
binning. Effectively, an overall better agreement in some distributions like $m_{\gamma\gamma}$ and 
at large 
$q_T$ is traded for a worse agreement at small $q_T$, $\phi^*$ \cite{Banfi:2010cf} or $a_T$ 
\cite{Vesterinen:2008hx} as well 
as for azimuthal photon separations
$\Delta\Phi\sim \pi$. While these are all regions that need to be addressed by the resummation of 
large $q_T/Q$ logarithms, we show that the discontinuity effects are amplified and that the 
agreement with data is \emph{considerably} worsened: For the $q_T$ distribution around 
$E_T^\text{iso}$ and equivalently 
for $\Phi^*$, the agreement of central values within a few percent is turned into disagreement of 
50-60\%, see our 
results in the 
following. Indeed in 
ref.~\cite{Gehrmann:2020oec} the authors anticipated problems with 
slicing subtractions. The $q_T$ resummation (at leading power) in that sense acts like a slicing 
procedure, 
adding only 
Born-topology corrections on top of the fixed-order prediction that exhibits the Sudakov 
singularity. There is no compensating mechanism from the fixed-order expansion of the resummed 
result, resulting in the large unphysical matching corrections. The hybrid-cone isolation and 
natural scale choice also can unfortunately not help to address the large difference between 
$\alpha_s$ and $\alpha_s^2$ results, but they raise valid concerns about previous assumptions.

Exactly this current situation makes a calculation of higher-order ($\alpha_s^3$) effects 
necessary. They can 
	hopefully unambiguously stabilize the perturbative series to allow for truncation uncertainties 
	that can be trusted by finding overlapping bands between different orders in both the large and 
	small $q_T$ regions.  A first step in that 
	direction at large $q_T$ is the very recent \NNLO{} calculation of $\gamma\gamma$+jet 
	\cite{Chawdhry:2021hkp}. This calculation predicts positive corrections at the order of 10\% 
	(without the loop-induced $gg$ channel)
	below \SI{100}{\GeV}, which likely fills the currently seen gaps between $\NLO{}$ 
	large-$q_T$ predictions and data, see e.g. fig.~18 in ref.~\cite{Becher:2020ugp}.
	Also, since using the hybrid-cone isolation scheme with large-$q_T$ \NLO{} predictions leads to 
	agreement with data within uncertainties, 
	see ref.~\cite{Gehrmann:2020oec} and our plots in the following,  it will be interesting to 
	investigate how 
	the additional 10\% effects from a large-$q_T$ \NNLO{} prediction behave in the presence of 
	this 
	isolation 
	scheme.

	In the present study we demonstrate the effect of the $\alpha_s^3$ hard functions, which are 
	three-loop for the $q\bar{q}$ \cite{Caola:2020dfu} channel and two-loop for the $gg$ channel 
	\cite{Bern:2001df} (implemented in \MCFM{} in refs.~\cite{Campbell:2016yrh,Campbell:2011bn}),
	and incorporate them with the recently published three-loop beam functions
	\cite{Luo:2020epw,Ebert:2020yqt,Luo:2019szz}. These ingredients are commonly referred to as 
	\enquote{constant} pieces and including them to a  higher order constitutes the primed
	accuracy, i.e. designated as $\NNNLL{}^\prime+\NNLO{}_0$.
	The inclusion of the primed contributions is typically a dominant effect of the next order and 
	also
	largely responsible for stabilization of truncation uncertainties. This can be seen by 
	comparing for example resummed spectra at \NLL{}$'$ with \NNLL{} and \NNLL{}$'$ with \NNNLL{}. 
	It is particularly true when matching corrections are small, such as in Drell-Yan 
	production. This was recently observed in refs.~\cite{Re:2021con,Ju:2021lah}, 
	and we also observe it in our N$^3$LL$^\prime$ implementation for Drell-Yan.
	
	But while for Drell-Yan production $q_T$ resummation works up to 40--50$\,$GeV with 
	almost
	negligible 
	(1-2\%) 
	matching corrections \cite{Ebert:2020dfc,Becher:2020ugp}, the situation for photon processes is 
	different due 
	to the photon isolation \cite{Ebert:2019zkb}. 
	With fiducial cuts the photon isolation prescription induces large linear power corrections 
	\cite{Becher:2020ugp}. 
	Furthermore, the typical minimum $q_T^{\gamma,1}$ and 
	$q_T^{\gamma,2}$ cuts on the two photons completely invalidate $q_T$ resummation above $\sim 
	q_T^{\gamma,1}+q_T^{\gamma,2}$ ($=\SI{70}{\GeV}$ for the \ATLAS{} \SI{8}{\TeV} study) and the 
	matching corrections quickly grow towards that point. With these effects taken together, 
	the matching corrections 
	from fixed-order are 
	50-75\% over the whole range of applicable $q_T$. This means that more than half of the 
	cross-section at 
	small $q_T$ comes from the terms of the fixed-order prediction and are not 
	described by the higher-order $q_T$-logarithms.	
	 Ideally one would like to resum the linear power corrections 
	$\mathcal{O}(q_T/Q)$, but the isolation makes this difficult.

	While several effects diminish stabilizing effects from the inclusion of the three-loop 
	$q\bar{q}$ hard function, the present calculation allows for a first inclusion of the 
	three-loop virtual corrections in a physical calculation without the complication of a full 
	N$^3$LO calculation, and therefore shows directly the impact of including these corrections.
	We also treat for the first time the $q\bar{q}$ and $gg$ loop-induced $\alpha_s^3$ hard 
	functions fully consistently together in the resummation. Both channels have to be added 
	separately together, of course, which means to reach the $\alpha_s^3$ accuracy for the 
	\enquote{constant} part, we add the \NNNLL{}$^\prime$ resummed $q\bar{q}$ channel to the 
	\NNLL{} 
	resummed $gg$ loop-induced channel. 
	
	\paragraph{Implementation.} We extend the existing framework 
	\CuTeMCFM{} \cite{Becher:2020ugp} which implements \NNNLL{} $q_T$ resummation in the 
	\SCET{} formulation of 
	refs.~\cite{Becher:2010tm,Becher:2011xn} matched to fixed-order 
	$\NNLO{}_0$ \cite{Campbell:2011bn,Campbell:2016yrh}. This framework achieves an 
	accuracy 
	of $\alpha_s^2$	in improved perturbation theory at small and large $q_T$.
	To upgrade to \NNNLL{}$^\prime$ accuracy we have implemented the one-, two- and three-loop 
	$\overline{\text{MS}}$-renormalized virtual amplitudes from 
	ref.~\cite{Caola:2020dfu}  and restored the 
	renormalization-scale dependence by solving the associated \RGE{} \cite{Becher:2009qa} to order 
	$\alpha_s^3$. For the numerical evaluation of harmonic polylogarithms up 
	to weight six in the hard function we use 
	the \texttt{hplog} library 
	\cite{Gehrmann:2001pz}.\footnote{We would like to thank Thomas Gehrmann for providing us with a 
	version that computes the harmonic polylogarithms up to weight six.} 
	After that we find full agreement to machine 
	precision with the existing one- and 
	two-loop results in \MCFM{}. The implemented $\overline{\text{MS}}$-renormalized amplitudes 
	constitute the hard function. We also implemented the three-loop 
	beam functions \cite{Luo:2020epw,Ebert:2020yqt,Luo:2019szz} and find that the double 
	logarithmic $L_\perp\sim\log(x_T^2\mu^2)$-dependence, where $x_T$ is the fourier-conjugate of 
	$q_T$, is as predicted by associated 
	renormalization group equations \cite{Becher:2010tm}. For the resummation we employ an improved 
	power counting 
	$L_\perp\sim 1/\sqrt{\alpha_s}$ (relevant 
	at small $q_T$), factor out the double-logarithmic $L_\perp$-dependence of the beam functions 
	and exponentiate it through associated \RGE{} \cite{Becher:2010tm}. In addition to the 
	previously published version of \CuTeMCFM{} \cite{Becher:2020ugp} we have also made the 
	resummation scale uncertainties more robust by additionally varying the rapidity scale 
	following ref.~\cite{Jaiswal:2015nka}.
	
	\paragraph{Results.}
	
	Before showing differential results, we first discuss fiducial total 
	cross-sections. At fixed order we can calculate these up to \NNLO{}. For the $q_T$-resummed 
	predictions we can simply integrate over $q_T$ to obtain a total cross-section which includes
	higher-order logarithmic corrections. Ideally this is within the scale-uncertainties of the 
	fixed-order result. Since the bulk of the cross-section comes from small $q_T$, the resummation 
	can, in principle, improve the prediction and uncertainties. With \NNNLL{}$^\prime$ $q_T$ 
	resummation 
	we can give a consistent prediction including both the $q\bar{q}$- and $gg$-initiated hard 
	functions at 	order  $\alpha_s^3$.
	
	For all results which follow, we implemented the parameter choices and cuts from the \ATLAS{} 
	\SI{8}{\TeV} 
	study in 
	ref.~\cite{Aaboud:2017vol}. The selection cuts are $q_T^{\gamma,\text{hard}}>\SI{40}{\GeV}$, 
	$q_T^{\gamma,\text{soft}}>\SI{30}{\GeV}$,	$\abs{\eta_\gamma}<2.37$, omitting 
	$1.37<\abs{\eta_\gamma}<1.56$, $R_{\gamma\gamma}>0.4$. The 
	smooth-cone photon isolation  criterion  is used with 
	$E_T^{\text{iso}}=\SI{11}{\GeV}$,  $n=1$ and $R_s=0.4$. Throughout we use the 
	{\texttt{NNPDF31\_nnlo\_as\_0118}} \PDF{} set \cite{NNPDF:2017mvq}.  The 
	measured fiducial cross-section is $\SI{16.8\pm0.8}{pb}$.
	
	In \cref{tab:totcross8tevqq} we present the cross-sections for the $q\bar{q}$ and $gg$
	hard-function initiated processes at fixed-order and by integrating resummed cross-sections 
	over $q_T$. Our resummation is matched to fixed-order predictions using a transition function 
	as detailed in ref.~\cite{Becher:2020ugp}.
	The matching uncertainty from varying the transition function is about one percent at higher 
	orders, 
	i.e. small compared to the other uncertainties. We neglect matching corrections below 
	\SI{1}{\GeV}, which has an effect smaller than the numerical precision quoted.
	  For the fixed-order 
	cross-sections the numbers in front of $\mu_R$ and $\mu_F$ denote a variation of the 
	renormalization and factorization scale by a factor of $2$ and $1/2$, respectively. We only 
	quote the maximum of upwards and downwards variation and take these as symmetric uncertainties. 
	We also 
	vary combinations, but their uncertainties are found to be smaller or similar. For the resummed 
	predictions we correlate the renormalization scale $\mu_R$ with the hard scale and correlate 
	the resummation scale with 	the 
	factorization scale $\mu_F$ as in ref.~\cite{Becher:2020ugp}. To obtain robust uncertainties we 
	additionally vary the rapidity-\RGE{} scale, but find that these 
	uncertainties are smaller than the resummation scale uncertainties for this process. In most 
	studies all of these uncertainties are combined by taking the envelope. Here we just quote both 
	numbers.
	
\begin{table}
	\caption{Cross-sections at \SI{8}{\TeV} for $q\bar{q}$ and $gg$ hard-function initiated 
	processes with $\mu_R=m_{\gamma\gamma}$ and smooth-cone isolation. }
	\label{tab:totcross8tevqq}
	\centering
	\begin{tabular}{@{}c|c@{}}
		\toprule
		& $\sigma_{\bar{q}q} / \text{pb}$ \\ \midrule
		\NLO{} &    $9.7\pm0.7(\mu_R)\pm0.4(\mu_F)$    \\
		
		$\int \NNLL{}+\NLO{}_0$ & $8.9\pm1.1(\mu_R)\pm0.9(\mu_F)$     \\ \midrule
		
		
		\NNLO{} & $12.8\pm1.8(\mu_R)\pm0.2(\mu_F)$  \\
		
		$\int \NNNLL{}+\NNLO{}_0$&	$12.7\pm0.9(\mu_R)\pm0.4(\mu_F)$ \\ \midrule
		
		$\int \NNNLL{}^\prime+\NNLO{}_0$& $12.8\pm0.9(\mu_R)\pm0.4(\mu_F)$  \\ \bottomrule
	\end{tabular}

	\vspace{1.0em}
	
	\begin{tabular}{@{}c|c@{}}
	\toprule
	& $\sigma_{gg} / \text{pb}$ \\ \midrule
	\LO{} &  $1.0\pm0.2(\mu_R)\pm0.0(\mu_F)$   \\ \midrule
	\NLO{} & $2.0\pm0.4(\mu_R)\pm0.1(\mu_F)$  \\
	$\int \NNLL{}+\NLO{}_0$ & $1.0\pm0.5(\mu_R)\pm0.1(\mu_F)$     \\ \bottomrule
\end{tabular}
\end{table}

At \NLO{} and \NNLO{} we find indeed that the integrated resummed predictions agree within 
uncertainties with the fixed-order result. At order $\alpha_s^2$ (\NNLO{}) the uncertainties of the 
resummed prediction are smaller.
 The primed prediction taking into account the 
$\alpha_s^3$ constant 
terms ($\int \NNNLL{}^\prime+\NNLO{}_0$) shows no significant changes compared to the $\alpha_s^2$ 
prediction, neither in cross-section nor in uncertainties. To understand that underlying this is 
still 
a 
stabilization of the cross-section we have to look at the results differentially next. 
Taking together the $q\bar{q}$ and $gg$ contributions, the prediction and measurement agree at best 
marginally within mutual uncertainties, which is a known issue.

As suggested in ref.~\cite{Gehrmann:2020oec} a more 
natural choice for the renormalization (hard) and factorization scale is to take the arithmetic 
average of the photon 
momenta $\mu_R=\langle q_T^\gamma\rangle$, so we consider this in addition to our standard 
choice of $\mu_R=\sqrt{(m_{\gamma\gamma})^2+(q_T^{\gamma\gamma})^2}$. With this scale we present 
results 
at the highest orders in \cref{tab:totcross2}. The cross-section increases noticeably by more than 
the size of the previous scale variation uncertainty, and is now in much better agreement with the 
measurement.

\begin{table}
	\caption{Cross-sections at \SI{8}{\TeV} for $q\bar{q}$ and $gg$ hard-function initiated 
		processes with $\mu_R=\langle q_T^\gamma \rangle$ and smooth-cone isolation. }
	\label{tab:totcross2}
	\centering
	\begin{tabular}{@{}c|c@{}}
		\toprule
		& $\sigma_{\bar{q}q} / \text{pb}$ \\ \midrule
		$\int \NNNLL{}^\prime+\NNLO{}_0$&  $13.7\pm1.0(\mu_R)\pm0.3(\mu_F)$   \\ \bottomrule
	\end{tabular}
	
	\vspace{1.0em}
	
	\begin{tabular}{@{}c|c@{}}
		\toprule
		& $\sigma_{gg} / \text{pb}$ \\ \midrule
		$\int \NNLL{}+\NLO{}_0$ &  $1.8\pm1.0(\mu_R)\pm0.2(\mu_F)$  \\ \bottomrule
	\end{tabular}
\end{table}

\begin{figure}
	\includegraphics[width=\columnwidth]{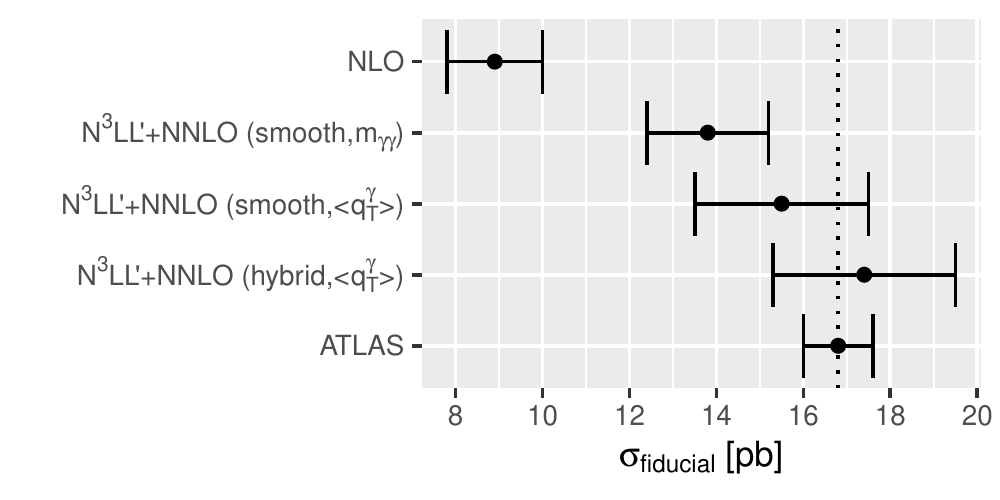}
	\caption{Comparison of selected fiducial cross-sections from 
	\cref{tab:totcross8tevqq,tab:totcross2,tab:totcross3} with the \SI{8}{\TeV} \ATLAS{} 
	measurement. Here both 
	the 
	$q\bar{q}$ and loop-induced $gg$ channel contributions are added together consistently. }
	\label{fig:fiducial}
\end{figure}

Additionally we compare with a hybrid-cone implementation where the inner smooth-cone radius is 
reduced to $R_s=0.1$ and the outer radius is $R_o=0.4$. The value of $R_s=0.1$ has been suggested 
in 
ref.~\cite{Gehrmann:2020oec} to best match the \ATLAS{} \SI{8}{\TeV} measurement and 
the fragmentation calculation.
We therefore present results with both $\mu_R=\langle q_T^\gamma \rangle$ and an inner cone 
$R_s=0.1$ in 
\cref{tab:totcross3}. This time the central value already overshoots the measured central value, 
but is also within uncertainties. To visualize the results of 
\cref{tab:totcross8tevqq,tab:totcross2,tab:totcross3} and directly 
compare with the measurement, we show selected combinations in \cref{fig:fiducial}, where the 
$q\bar{q}$ 
and 
loop-induced $gg$ channel are combined consistently at the same order of $\alpha_s$. 
\begin{table}
	\caption{Cross-sections at \SI{8}{\TeV} for $q\bar{q}$ and $gg$ hard-function initiated 
		processes with $\mu_R=\langle q_T^\gamma \rangle$ and hybrid-cone isolation with inner 
		cone angle $R_0=0.1$. }
	\label{tab:totcross3}
	\centering
	\begin{tabular}{@{}c|c@{}}
		\toprule
		& $\sigma_{\bar{q}q} / \text{pb}$ \\ \midrule
		$\int \NNNLL{}^\prime+\NNLO{}_0$&  $15.5\pm1.1(\mu_R)\pm0.4(\mu_F)$   \\ \bottomrule
	\end{tabular}
	
	\vspace{1.0em}
	
	\begin{tabular}{@{}c|c@{}}
		\toprule
		& $\sigma_{gg} / \text{pb}$ \\ \midrule
		$\int \NNLL{}+\NLO{}_0$ &  $1.9\pm1.0(\mu_R)\pm0.2(\mu_F)$  \\ \bottomrule
	\end{tabular}
\end{table}

	\paragraph{The resummed $q_T$ spectrum without matching.}
	
\begin{figure}[h]
		\includegraphics[width=\columnwidth]{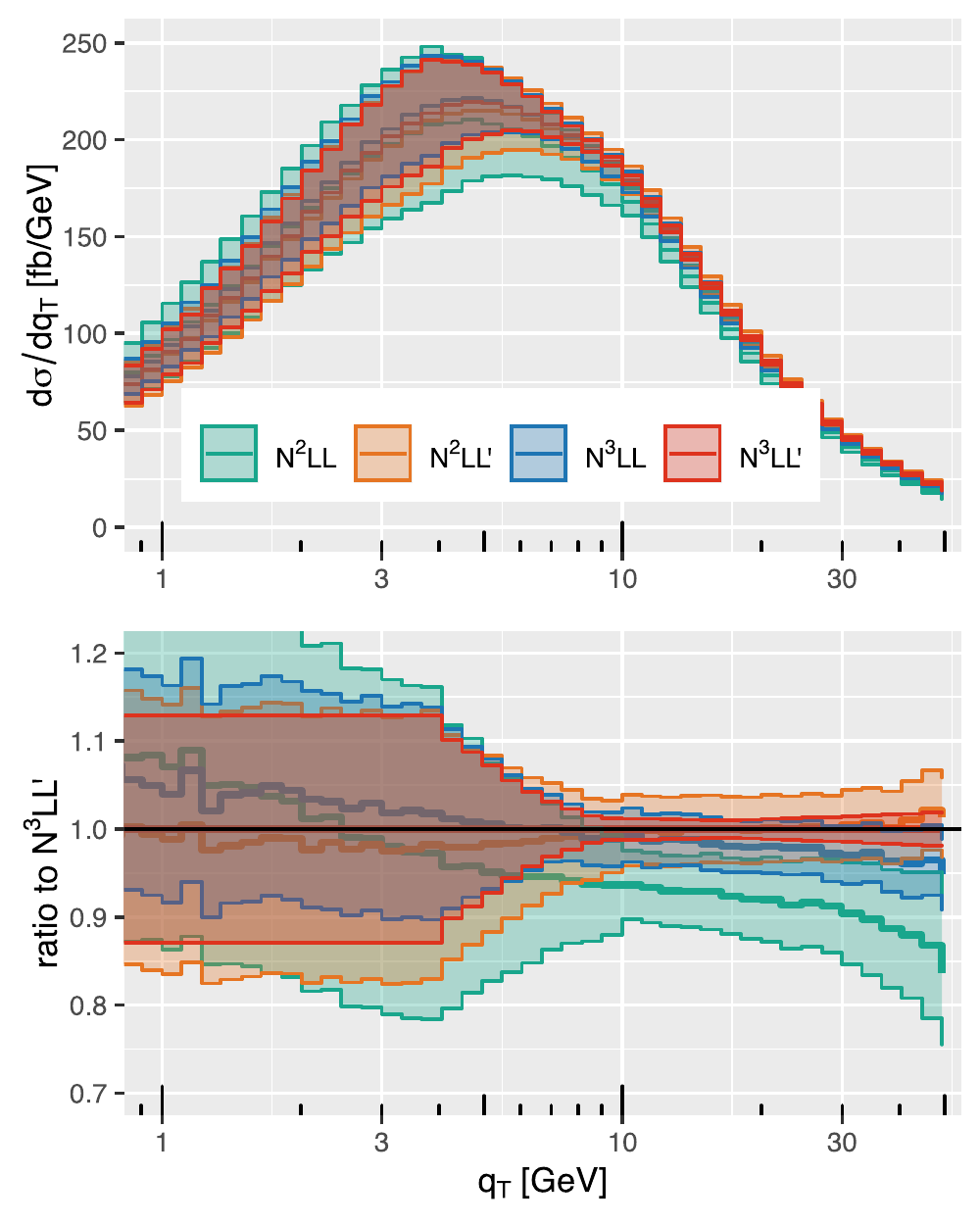}
		\caption{Resummed $q_T$ spectrum without transition function nor matching for 
		$q\bar{q}$-initiated hard function at different orders. Top plot: 
		absolute distributions. Bottom plot: ratio to \NNNLL{}$^\prime$.}
		\label{fig:purelyresummed}
	\end{figure}

	Differentially, we first discuss the resummed $q_T$ distributions without matching. We also 
	focus only on the $q\bar{q}$ channel 
	first, where 
	we now include 
	the $\alpha_s^3$ hard function at \NNNLL{}$^\prime$. This is shown in \cref{fig:purelyresummed}.

	 The bottom plot shows the ratio to the highest-order prediction \NNNLL{}$^\prime$, while the 
	 upper 
	 plot shows the absolute distributions. The \NNLL{} result takes into account just the 
	 $\alpha_s$ hard function, while \NNLL{}$^\prime$ and \NNNLL{} take into account the 
	 $\alpha_s^2$ hard 
	 function but differ in the order of \RGE{} solution that resums the large logarithms. Only 
	 \NNNLL{} is fully consistent to order $\alpha_s^2$ in improved perturbation theory.
	 
	 The displayed uncertainties are obtained from the envelope of a variation of hard scale, 
	 resummation scale and rapidity scale. Since we use the envelope, the largest 
	 uncertainty determines the band, which in our case is from the resummation scale. Below 
	 \SI{4}{\GeV} the uncertainties turn constant for 
	 the following reason. In any resummation formalism a cutoff at small $q_T$ is 
	 necessary since, for example, otherwise $\alpha_s$ would be evaluated at scales where 
	 non-perturbative  effects become significant, i.e. where $\alpha_s$ becomes large.
	 We choose to set a minimum scale of \SI{2}{\GeV}, 
	 which consequently leads to frozen out uncertainties below \SI{4}{\GeV} when a 
	 downwards scale variation becomes ineffective. Without such a 
	 cutoff the uncertainties would become arbitrarily large and would not represent realistic 
	 perturbative uncertainties, in addition to numerical problems.
	 
	 Overall the  uncertainties decrease going from \NNLL{} to higher orders. At the smallest $q_T$ 
	 the uncertainties for \NNLL{} are about 20\% and reduce to 12\% going towards \NNNLL{}, but 
	 change little between \NNNLL{} and \NNNLL{}$^\prime$, likely an effect due to the same order 
	 in \RGE{} 
	 running. For larger $q_T$ we compare \NNLL{}, \NNLL{}$^\prime$ and \NNNLL{} and find that the 
	 primed 
	 accuracy, taking into account the higher-order hard 
	 and beam functions but solving the \RGE{}s to a lower order, is responsible for the bulk of 
	 corrections. This is reflected by the good agreement between the orange and blue lines. 
	 Consequently one expects that the inclusion of the $\alpha_s^3$ hard and beam functions 
	 is responsible for the bulk of corrections within a consistent N$^4$LL calculation. Indeed the 
	 highest-order prediction \NNNLL{}$^\prime$ is between both lower order predictions and 
	 noticeably 
	 decreases uncertainties above \SI{10}{\GeV}.
	 
	 \begin{figure}[h]
	 	\includegraphics[width=\columnwidth]{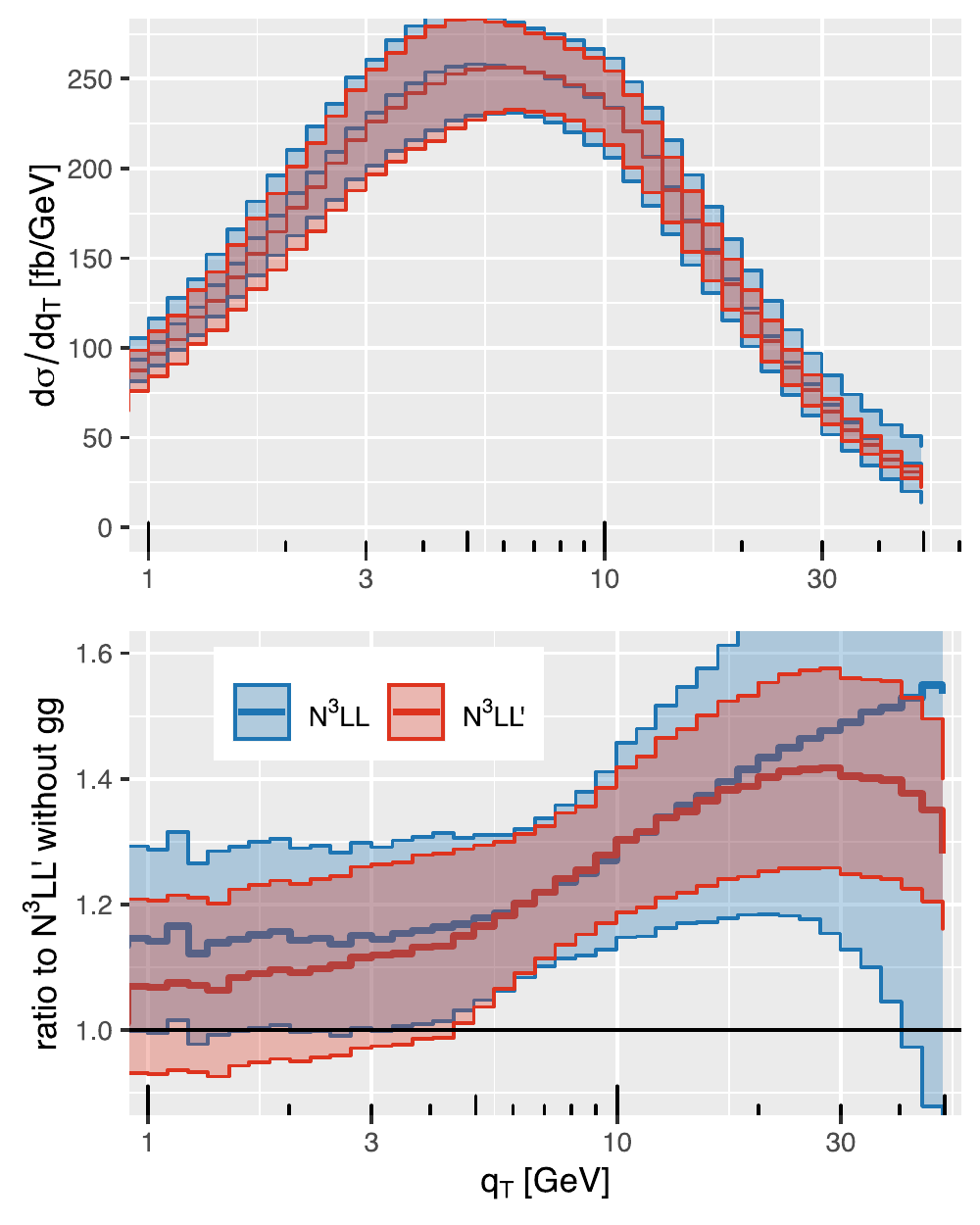}
	 	\caption{Resummed $q_T$ spectrum without transition function nor matching including both 
	 	$q\bar{q}$ and $gg$ channels. Top plot: 
	 		absolute distributions. Bottom plot: ratio to \NNNLL{}$^\prime$ without $gg$ channel.}
	 	\label{fig:purelyresummedwithgg}
	 \end{figure}	
	 
	 These finding indicate a stabilization of the $q\bar{q}$ channel, but at order 
	 $\alpha_s^2$ the loop-induced $gg\to\gamma\gamma$ channel enters that is enhanced due to the 
	 large 
	 gluon 
	 luminosity at the \LHC{} at small momentum fractions. In \cref{fig:purelyresummedwithgg} we 
	 include this channel at the respective orders, i.e. the $\alpha_s^2$ hard function at \NNNLL{} 
	 and  the  $\alpha_s^3$ hard function at \NNNLL{}$^\prime$. The $gg$ channel is a substantial 
	 contribution with \emph{huge} uncertainties. At low $q_T$ the uncertainties are still at the 
	 order of 10\%. Towards larger $q_T$ the $gg$ channel contributes \enquote{only} half of the 
	 cross-section, but the uncertainties are so large that the $q\bar{q}$ channel uncertainties of 
	 1--2\% seen in \cref{fig:purelyresummed} blow up to 10\% in the sum of both channels. This is 
	 not unexpected since the $gg$ channel, despite being of order $\alpha_s^3$ is only \NLO{}, 
	 respectively \NNLL{}+\NLO{} accurate. To increase the precision in the intermediate low $q_T$ 
	 region of about 10-50~GeV where resummation remains relevant, three-loop 
	 $\alpha_s^4$ corrections to the $gg$ channel will therefore be important.

	\paragraph{Fully matched results.}
	
	We now move on to show fully matched results and directly compare with the \SI{8}{\TeV} 
	\ATLAS{} 
	measurement. Our transition function is a function of $x=q_T^2/m_{\gamma\gamma}^2$ with a 
	parameter $x^\text{max}$ that determines the transition region, see ref.~\cite{Becher:2020ugp} 
	for a detailed description.	For the following plots we use $x^\text{max}=0.1$ that performs 
	the transition mostly in the region between $20$ and \SI{50}{\GeV}.  
	Since $m_{\gamma\gamma}$ is not sharply peaked 
	as in resonant boson production, 
	there is a tail of larger $m_{\gamma\gamma}$, for which the transition is later, such that we 
	need to choose $x^\text{max}$ relatively small to prevent reaching $\sim\SI{70}{\GeV}$	where 
	the 
	resummation 
	breaks down due to the given photon cuts. We estimate the matching uncertainty by varying the 
	transition function to use $x^\text{max}=0.2$. This shifts the 
	transition to be between $30$ and \SI{70}{\GeV}. The resulting difference is small compared to 
	our presented uncertainty bands obtained by scale variation. 
	
\begin{figure}[h]
		\includegraphics[width=\columnwidth]{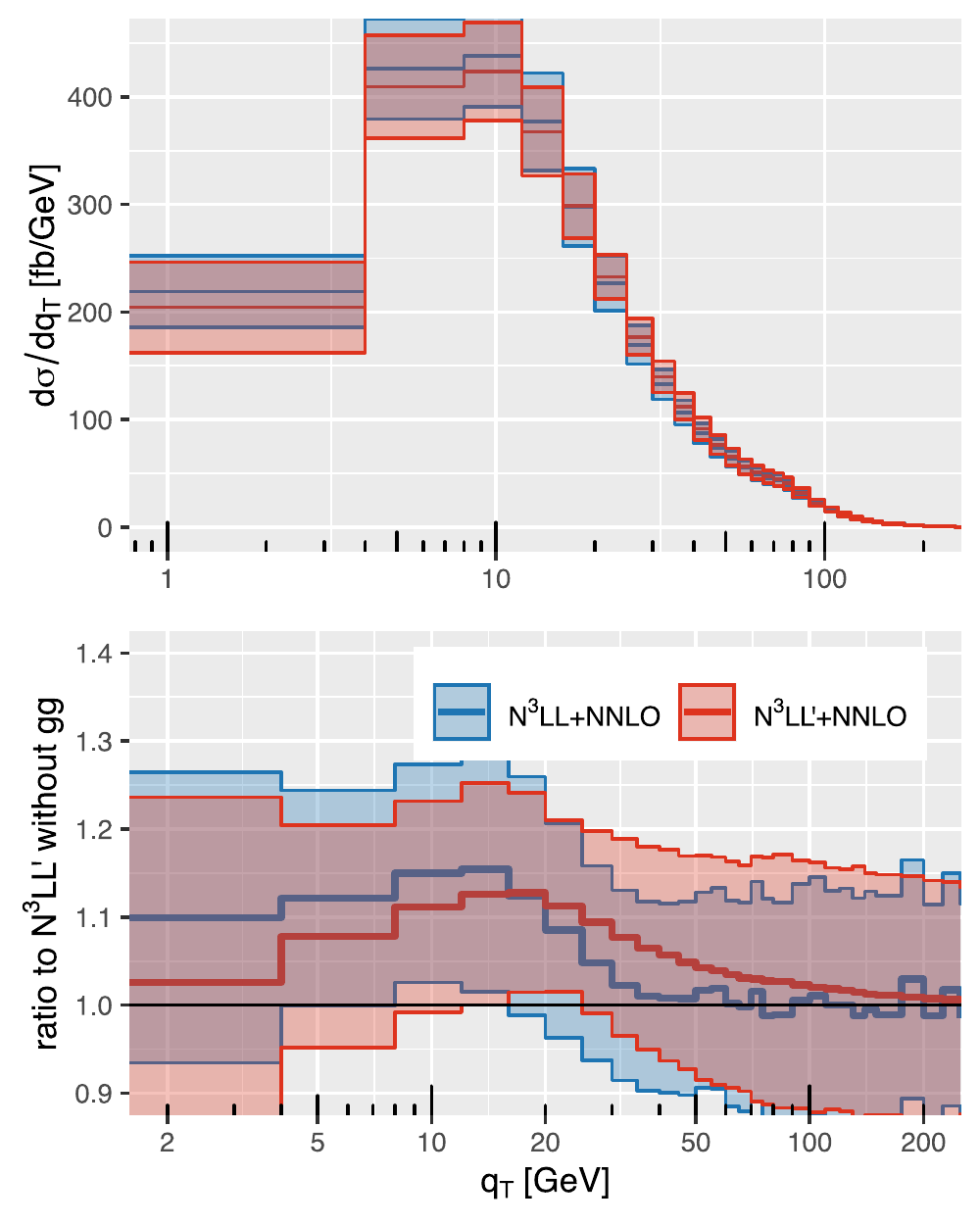}
		\caption{Fully matched $q_T$ spectra at \NNNLL{}+\NNLO{} and \NNNLL{}$^\prime$+\NNLO{}. Top 
		plot: 
			absolute distributions. Bottom plot: ratio to \NNNLL{}$^\prime$+\NNLO{} without $gg$ 
			channel.}
		\label{fig:matched_ggratio}
	\end{figure}

	So far the stabilization of the resummed $q\bar{q}$ channel has been somewhat overshadowed by 
	large 
	$gg$ channel uncertainties. We now show the fully matched result in \cref{fig:matched_ggratio} 
	with \ATLAS{} binning. The matching corrections at $\alpha_s^2$ for diphoton production are 
	sizable about 50\%, as can also be seen by comparing with \cref{fig:purelyresummedwithgg}. The 
	top plot shows the absolute predictions for \NNNLL{}+\NNLO{} and \NNNLL{}$^\prime$+\NNLO{}, 
	while the 
	lower plot shows the ratio to the higher-order prediction without the $gg$ channel. The 
	higher-order corrections from the $q\bar{q}$ channel are small as we have seen, but the 
	$\alpha_s^3$ corrections on the $gg$ channel have a noticeable impact (at large $q_T$ we 
	include the $\alpha_s^3$ matching corrections in this channel). In both cases the uncertainties 
	are large and transition into fixed-order uncertainties of about 
	15\% at large $q_T$. Overall both predictions show uniform uncertainties of 10-15\%.

	To decrease uncertainties noticeably we will first have to include $\alpha_s^3$ 
	matching-corrections also in the $q\bar{q}$ channel, 
	which at low $q_T$ make up about 50\% of the cross-section. Second, the $gg$ channel has 
	to be included at $\alpha_s^4$ since it contributes an equal amount to the total uncertainty.
	
\begin{figure}[h]
	\includegraphics[width=\columnwidth]{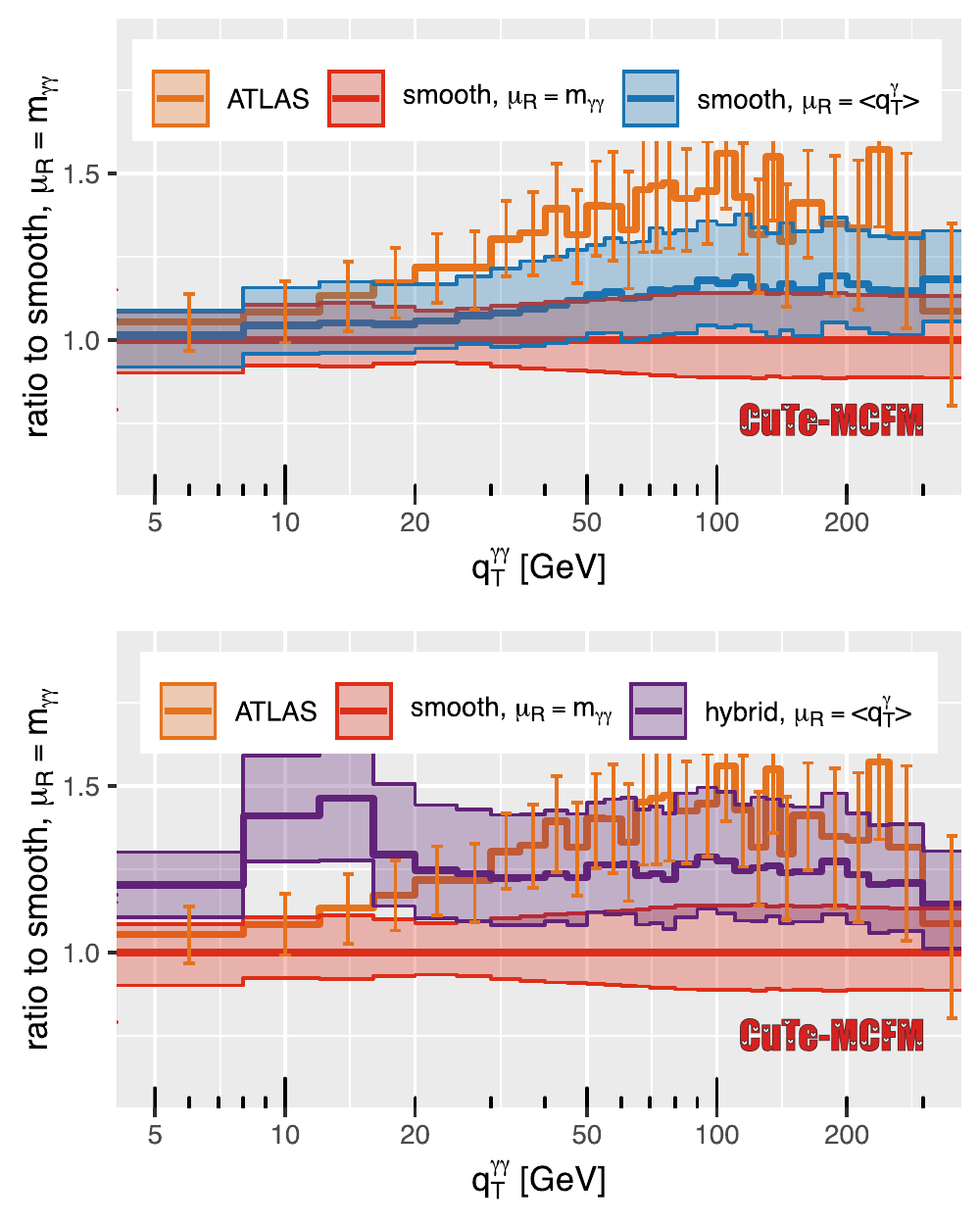}
	\caption{Fully matched $q_T$ spectra at \NNNLL{}$^\prime$+\NNLO{} in comparison with the 
	\ATLAS{} 
	measurement. Top 
	plot: Ratio to \NNNLL{}$^\prime$ with smooth-cone isolation and $\mu_R=m_{\gamma\gamma}$ in 
	comparison 
	with data and $\mu_R=\langle q_T^\gamma\rangle$. Bottom plot: Similar, but in comparison with 
	prediction using hybrid-cone isolation and $\mu_R=\langle q_T^\gamma\rangle$.  }
	\label{fig:matched_expratio}
\end{figure}

	We finally show the fully matched results in comparison with the \ATLAS{} measurement in 
	\cref{fig:matched_expratio}. The top plot shows the ratio of the \ATLAS{} measurement to our 
	highest-order prediction as in \cref{fig:matched_ggratio}. We furthermore included a prediction 
	where the hard scale is chosen as $\langle q_T^\gamma \rangle$ as suggested in 
	ref.~\cite{Gehrmann:2020oec}. This more natural scale choice closes the uncertainty gap, and 
	prediction and measurement have now overlapping uncertainty bands.
	
	The bottom plot includes additionally a prediction with the hybrid-cone isolation using an 
	inner-cone radius of $R_s=0.1$, where the previous agreement at small $q_T$ is now destroyed.
	As already shown in ref.~\cite{Gehrmann:2020oec}, the fixed-order predictions with hybrid-cone 
	isolation at small $\Phi^*$ are worse than the smooth-cone isolation results.\footnote{To our 
	surprise they do not find this to be true for small $q_T$, even though the region of small 
	$q_T$ and $\Phi^*$ should map onto each other.} The authors suggest that \enquote{The regions 
	where agreement is notably 
	worse are those in the neighborhood of the Sudakov singularities [...], and hence where poor 
	agreement is expected in the absence of resummation}. While this is true, the distortion due to 
	the hybrid-cone isolation of the $q_T$ distribution cannot be cured by the present $q_T$ 
	resummation, but would require some other resummation. 
	
	In the limit of inner cone approaching outer cone $R_s\to R_o$ the smooth-cone isolation is 
	restored by definition. In this limit the resummed results agree well with the data. For 
	successively smaller $R_s$ the cross-section coming from fixed-cone 
	isolation grows. Since it is always larger than the contribution from smooth-cone isolation, 
	the agreement at large $q_T$ also grows. But the 
	increase in cross-section is unfortunately not just at large $q_T$: the smaller $R_s$ is taken, 
	the larger the ridge at $q_T=E_T^\text{iso}$ becomes (see e.g. fig.~8 in 
	ref.~\cite{Gehrmann:2020oec}). This ridge effect can 
	be smoothened out to some extend by choosing a non-constant $E_T^\text{iso}$, but would also 
	have 
	to be matched by the experimental definition.
	
	We conclude that with present theory frameworks the hybrid-cone isolation is not the answer to 
	a better modeling of photon isolation, especially with increased experimental precision. 
	Distortion effects in the $q_T$ distribution due to hybrid-cone isolation are only 
	exacerbated by the $q_T$ resummation. Better agreement at large $q_T$ is traded 
	with drastic disagreement at small $q_T$, where excellent agreement 
	with 	resummation is achieved using the smooth-cone isolation. While it is possible to shift 
	the Sudakov singularity in phase space, we believe that at this point the 
	program for fragmentation functions will have to be revived. 

\begin{figure}[h]
		\includegraphics[width=\columnwidth]{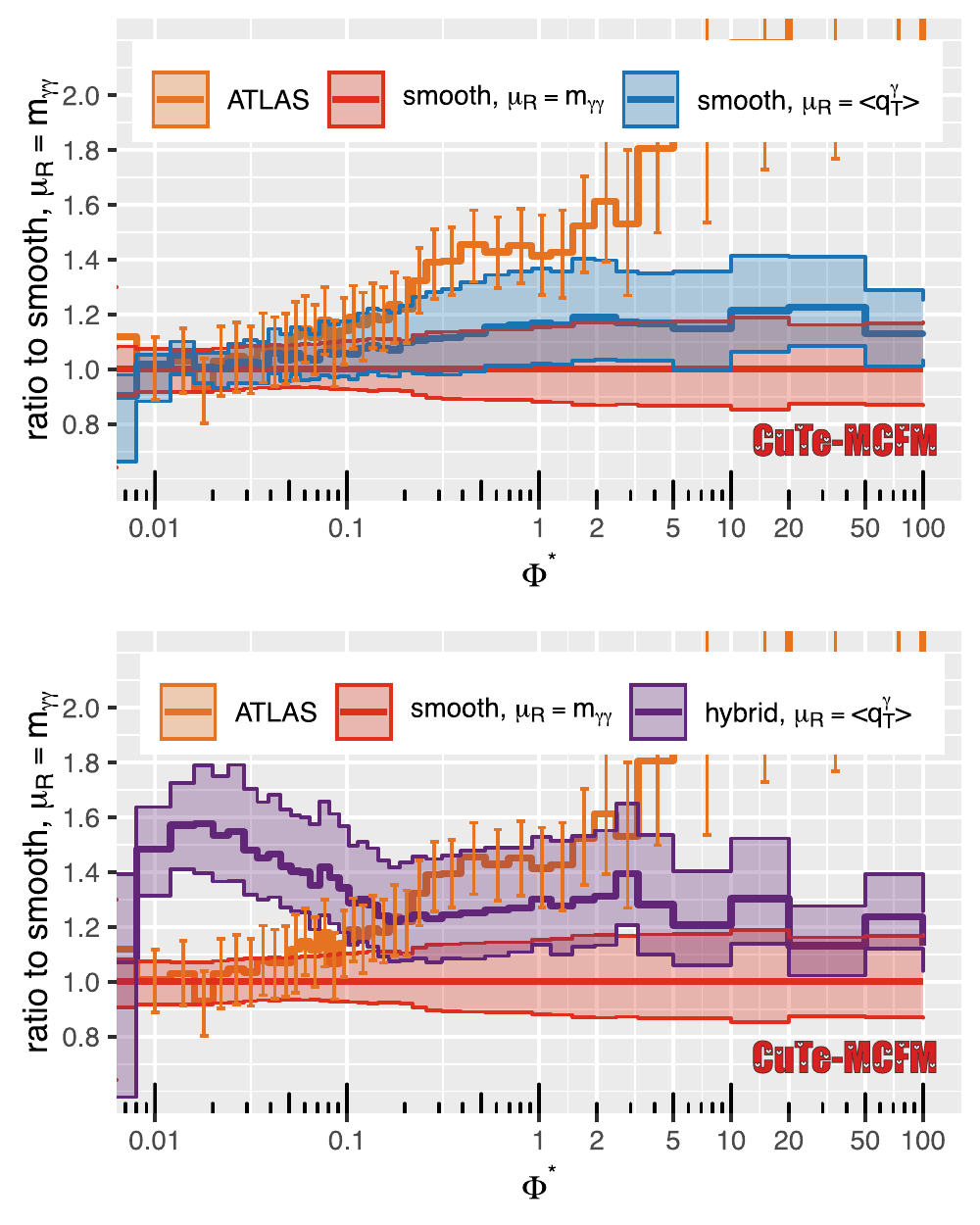}
		\caption{Fully matched $\phi^*$ spectra at \NNNLL{}$^\prime$+\NNLO{} in comparison with the
		\ATLAS{} 
		measurement. Top plot: Ratio to \NNNLL{}$^\prime$ with smooth-cone isolation and 
		$\mu_R=m_{\gamma\gamma}$ in comparison 
			with data and $\mu_R=\langle q_T^\gamma\rangle$. Bottom plot: Similar, but in 
			comparison with 
			prediction using hybrid-cone isolation and $\mu_R=\langle q_T^\gamma\rangle$.  }
		\label{fig:matched_expratio_phistar}
	\end{figure}

	For practical comparison with data, the lesson 
	to be learned is likely to just take the more natural scale choice $\mu_R=\langle 
	q_T^\gamma\rangle$, which brings theory and data into better agreement, and include the \NNLO{} 
	$\gamma\gamma$+jet corrections \cite{Chawdhry:2021hkp}. 
	We also show comparison plots for the 
	$\Phi^*$ distribution in \cref{fig:matched_expratio_phistar} with similar observations at small 
	$\Phi^*$, since it is directly correlated to small $q_T$, but with worse agreement at large 
	$\Phi^*$.	
		
	\paragraph{Conclusions.}
	
	We have upgraded previous diphoton predictions at small $q_T$ accurate at the level of 
	$\alpha_s^2$ in 
	improved perturbation theory to include the $\alpha_s^3$ \enquote{constant} pieces. These 
	include the recently published three-loop $q\bar{q}$ hard function \cite{Caola:2020dfu} and the 
	previously implemented two-loop $gg$ hard function \cite{Bern:2001df} together with the 
	three-loop transverse momentum dependent
	beam 
	functions \cite{Luo:2020epw,Ebert:2020yqt,Luo:2019szz}. This constitutes an overall primed 
	resummation accuracy of 
	$\NNNLL{}^\prime+\NNLO{}_0$.
	The resummation itself of the $q\bar{q}$ channel is noticeably stabilized with remaining  
	uncertainties of a few percent at intermediate $q_T$ between \SI{10}{\GeV} and \SI{50}{\GeV}. 
	But this is diminished by the large uncertainties from the $gg$ channel at $\alpha_s^3$ and 
	large matching corrections. 	
	
	With that, we have eliminated the $\alpha_s^3$ $q\bar{q}$ hard part as a source of uncertainty, 
	and can 
	limit the remaining sources of higher-order uncertainty and contributions for the $q_T$ 
	distribution: The dominating uncertainties will be reduced by a matching to 
	$\gamma\gamma$+jet 
	at \NNLO{}, which has ~10\% effects due to the $q\bar{q}$ channel below \SI{100}{\GeV} 
	\cite{Chawdhry:2021hkp}, and by
	$\alpha_s^4$ corrections to the $gg$ loop-induced channel, which are mostly relevant for 
	$q_T\lesssim\SI{200}{\GeV}$.
	
	We have shown that the more natural hard and renormalization scale $\langle q_T^\gamma 
	\rangle$ with smooth-cone 
	isolation alone restores agreement with data for the $q_T$ 
	distribution and for not too large $\Phi^*$. On the other hand the hybrid-cone isolation (with 
	fixed $E_T^\text{iso}$) introduces a discontinuity in the $q_T$ distribution that destroys the 
	otherwise agreement with data at small $q_T$. This 
	is because the isolation is only a power-suppressed effect in the resummation. The
	leading-power resummation acts at the level of the Born-topology without isolation effects and 
	cannot compensate the Sudakov singularity induced by the hybrid-cone isolation at fixed-order 
	\cite{Gehrmann:2020oec}.
	The upgrades presented in this paper will be included in the upcoming release of 
	\CuTeMCFM{}\footnote{\CuTeMCFM{} is available at \url{https://mcfm.fnal.gov/}.}.

	In the future we plan to match with a $\gamma\gamma$+jet calculation at \NNLO{} to take into 
	account the full $\alpha_s^3$ matching corrections at as small $q_T$ as possible. If small 
	enough $q_T$ can be achieved numerically, one could also extract fixed-order \NNNLO{} 
	cross-sections using $q_T$ slicing, but this will require cutoffs lower than \SI{1}{\GeV}, 
	which is already numerically quite expensive at \NNLO{} \cite{Campbell:2016yrh}. The 
	implementation 
	of the three-loop beam functions also facilitates an implementation of a single-boson 
	$\NNNLL{}^\prime+\NNLO{}_1$ implementation in \MCFM{}.
	
	\paragraph{Acknowledgments.} We would like to thank Thomas Becher, John Campbell and Sally 
	Dawson for useful discussion and comments on this manuscript. Tobias Neumann is supported by 
	the United States Department of Energy under Grant Contract DE-SC0012704.  This work was 
	supported by resources provided by the Scientific Data and Computing Center (SDCC),
	a component of the Computational Science Initiative (CSI) at Brookhaven National Laboratory 
	(BNL).
	
	\appendix
	
	\bibliographystyle{JHEP}
	\bibliography{refs}
	
\end{document}